\documentclass[prd, groupedaddress, showpacs, showkeys, onecolumn, nofootinbib]{revtex4}
\usepackage{amsfonts, amssymb, amsmath, amsthm, amscd}
\usepackage{graphicx}
\usepackage{hyperref}
\usepackage{bm}
\usepackage{enumerate}
\usepackage{amsfonts}%
\usepackage{mathrsfs}     
\begin{document} 
\title{Linking the trans-Planckian and information loss problems in black hole physics}
\author{Stefano Liberati}
\email{\tt liberati@sissa.it}
\affiliation{SISSA, Via Beirut 2-4, 34014 Trieste, Italy and INFN sezione di Trieste}
\author{Lorenzo Sindoni}
\email{\tt sindoni@sissa.it}
\affiliation{SISSA, Via Beirut 2-4, 34014 Trieste, Italy and INFN sezione di Trieste}
\author{Sebastiano Sonego}
\email{\tt sebastiano.sonego@uniud.it}
\affiliation{Universit\`a degli Studi di Udine, Via delle Scienze 208, 33100 Udine, Italy}
\date{July 25, 2009; \LaTeX-ed: \today}
\bigskip
\begin{abstract}
\bigskip

The trans-Planckian and information loss problems are usually discussed in the literature as separate issues concerning the nature of Hawking radiation. Here we instead argue that they are intimately linked, and can be understood as ``two sides of the same coin'' once it is accepted that general relativity is an effective field theory. 

\end{abstract}
\def\e{{\mathrm e}}%
\def\g{{\mbox{\sl g}}}%
\def\Box{\nabla^2}%
\def\d{{\mathrm d}}%
\def\ie{{\em i.e.\/}}%
\def\eg{{\em e.g.\/}}%
\def\etc{{\em etc.\/}}%
\def\etal{{\em et al.\/}}%
\newcommand{\scri}{\mathscr{I}}
\renewcommand{\thefigure}{\arabic{figure}}
\def\HRULE{{\bigskip\hrule\bigskip}}%
\def\implies{{\Rightarrow}}
\maketitle

\section{Introduction}
\label{sec:intro}

Although more than thirty years have passed since Hawking's discovery~\cite{Hawking:1974rv, Hawking:1974sw} of the quantum radiation leading to black hole evaporation, this effect is still the subject of intense investigation and debate. Among the issues studied, the trans-Planckian and information loss problems play a special role, not only for the difficulties related to their resolution, but also for the vast amount of literature devoted to them, as well as for the new lines of research they stimulated to undertake.

However, it is interesting to note that their status is nowadays not identical in different scientific communities.  In particular, while the trans-Planckian problem has been studied mainly by the gravitation theory community (see, \eg,~\cite{Unruh:1980cg,Unruh:1994je,Brout:1995wp,Corley:1996ar,Barrabes:1998iw,Parentani:1999qv,Barrabes:2000fr,Parentani:2000ts,Parentani:2007mb,Unruh:2004zk}), the information loss paradox, being a challenge to unitary evolution for quantum field theory in curved spacetimes~\cite{Hawking:1976ra, Hawking:1982dj}, has been in recent years investigated more actively by the community of particle physicists and string theorists (with remarkable results about the preservation of unitarity at the quantum gravity level~\cite{Stephens:1993an, 'tHooft:1996tq, Maldacena:2001kr}).

The point we want to make here is that there is a strong evidence that the two problems might be deeply intertwined, and that they could both be rooted in the effective field theory nature of general relativity (or other theories of gravity).  We shall argue that the information loss problem cannot be solved without addressing explicitly the trans-Planckian problem:  If the information loss admits a resolution, in the sense that there is a unitary evolution during the entire process of black hole formation and evaporation, this resolution must address the trans-Planckian problem too.

In the next section we briefly restate the trans-Planckian and the information loss problems, devoting particular care to the underlying assumptions.  Then, we describe in section~\ref{sec:scenario} a possible way out for the information loss paradox.  In section~\ref{sec:scatterings} we show, examining the properties of gravitational scatterings in the semiclassical limit, that the trans-Planckian problem represents a direct obstruction to the resolution of the information loss paradox within the proposed framework.  This allows us to argue, in section~\ref{sec:link}, that the information loss paradox is actually just a manifestation of the breakdown of the effective field theory used to describe black holes even in regions of small curvature, originated precisely by the trans-Planckian problem.  Section~\ref{sec:conclusions} contains the conclusions and a few remarks about other possible resolutions of the problems.

\section{Hawking radiation and its problems}
\label{sec:problems}

A physical situation in which the trans-Planckian and information loss problems appear, is the one in which a matter configuration in otherwise empty space collapses to form a black hole.  The spacetime dynamics associated with this process affects the behaviour of quantum fields, leading in particular to particle production.  For simplicity, consider a massless scalar field and restrict the analysis to spherically symmetric solutions.  Modes associated with outgoing particles behave  on $\scri^+$ as $\exp\left(-\mathrm{i}\,\omega u\right)/r$, where $u$ is the affine null coordinate on $\scri^{+}$ and $\omega>0$.  Similarly, modes associated with incoming particles at $\scri^-$ have the asymptotic form $\exp\left(-\mathrm{i}\,\Omega v\right)/r$, where $v$ is the affine null coordinate on $\scri^{-}$ and $\Omega>0$.  If one assumes that there are no incoming particles on $\scri^-$, the number of outgoing particles on $\scri^+$ can be obtained by evaluating the Bogoliubov coefficients between these modes~\cite{Hawking:1974sw, Birrell:1982ix, Novikov:1989sz}.  A mode of frequency $\Omega$ on $\scri^-$ can be approximately regarded as a mode of $u$-dependent frequency $\omega(u,\Omega) = \dot{p}(u)\,\Omega $ on $\scri^+$, where $v=p(u)$ is the function that links  $u$ and $v$ for the same null ray, and a dot denotes differentiation with respect to $u$.\footnote{Of course, this formula just expresses the frequency shift undergone by a signal in travelling from $\scri^-$ to $\scri^+$.}  Roughly, such a mode is excited during collapse if the standard adiabatic condition $|\,\dot{\omega}|/\omega^2\ll 1$ does \emph{not\/} hold~\cite{Barcelo:2007yk}.  This happens for frequencies on $\scri^-$ smaller than $\Omega_0(u) \sim |\,\ddot{p}(u)|/\dot{p}(u)^2$, so particles on $\scri^+$ will have typical frequencies smaller than $\omega_0(u)\sim |\,\ddot{p}(u)\,|/\dot{p}(u)$.  Using now the relation 
\begin{eqnarray}
p(u)\approx v_{\rm H} - A\e^{-\kappa u}\;,
\label{standard-p}
\end{eqnarray}
that holds for large values of $u$ under the assumption that a horizon forms at $u=+\infty$, $v=v_{\rm H}$ (here $\kappa=1/(4M)$ is the surface gravity, and $A>0$ is a constant that depends on the details of collapse), one finds the standard result for Hawking radiation, $\omega_0\sim \kappa$.  A more accurate analysis~\cite{Hawking:1974sw, Birrell:1982ix, Novikov:1989sz} shows that on $\scri^+$ there is a thermal flux of particles at temperature $\kappa/2\pi$.
\vskip 1.5cm
\begin{figure}[htb]
\begin{center}
\includegraphics[width=8.5cm]{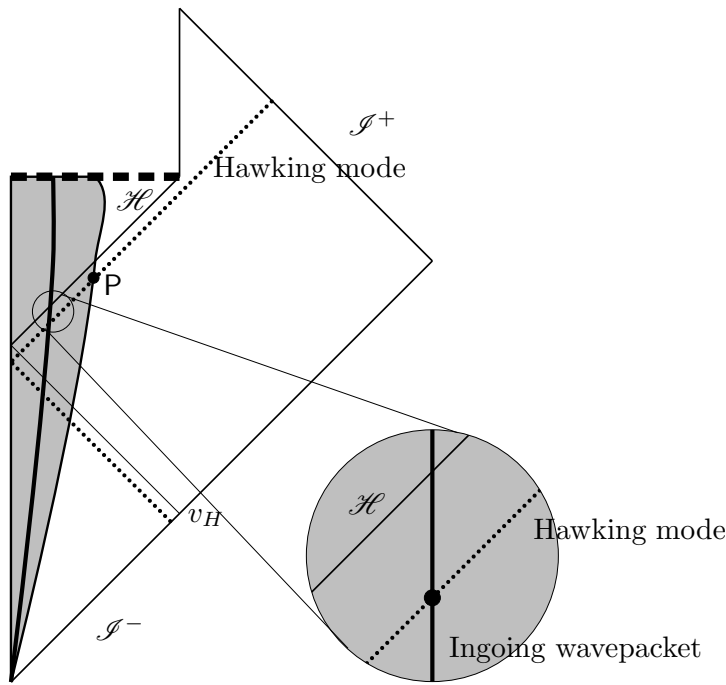}
\end{center}
\caption{Conformal diagram describing gravitational collapse followed by Hawking evaporation.  The detail shows the region in which a wavepacket in the collapsing configuration (solid thick line) interacts gravitationally with a Hawking mode (dotted line).}
\label{figure1}
\end{figure}
%

\subsection{Trans-Planckian problem}
\label{subsec:tP}

Hawking's result relies on the key assumption that quantum field theory continues to hold at arbitrarily high energies.  The typical frequency of the Hawking modes on $\scri^+$ is $\omega_0\sim\kappa$.  When traced back on $\scri^-$, these modes have frequency $\Omega_0(u)\sim\kappa/\dot{p}(u) \propto \e^{\kappa u}$, which for $u\rightarrow +\infty$ grows indefinitely.  Thus all the modes on $\scri^-$, even those that are trans-Planckian, are eventually excited.  This conclusion is problematic because it does not allow a decoupling between ultraviolet and infrared degrees of freedom.  Under the exponential relation~\eqref{standard-p} between $u$ and $v$, modes of arbitrarily low frequency on $\scri^+$ at late times contain arbitrarily high frequencies on $\scri^-$.\footnote{ In fact, in the physically realistic case of a black hole evaporating in a finite amount of time the frequencies excited by the Hawking process will not be arbitrarily high but  just well beyond the Planck scale.}  Hence, low energy physics on $\scri^+$ depends on the field behaviour at high energies on $\scri^-$, and no separation between a ``low energy'' and a ``high energy'' regime is applicable throughout the entire  spacetime.  

This trans-Planckian problem is unavoidable if an horizon forms (even only asymptotically~\cite{Barcelo:2006np, Barcelo:2006uw}).  Let $R$ be the Schwarzschild coordinate corresponding to the surface of the collapsing star, and let $\sf P$ be an event on that surface (see Fig.~\ref{figure1}).  $\sf P$ can be labeled by the null coordinates $u=t-R_\ast$ and $v=t+R_\ast$, where $R_\ast=R+2M\ln\left(R-2M\right)+\mbox{const}=\left(v-u\right)/2$ is Wheeler's tortoise coordinate.  In the limit $u\to +\infty$, $R\to 2M$, and $v$ tends to a finite value.  Hence, for large values of $u$, $(R-2M)\propto \mathrm{e}^{-\kappa u}\propto 1/\Omega_0(u)$. Therefore, modes of arbitrary frequency on $\scri^-$ are eventually excited when the surface of the star is sufficiently close to the location of the would-be horizon.  Hawking radiation detected on $\scri^+$ at late times is due to modes skimming the horizon arbitrarily closely, hence trans-Planckian on $\scri^-$.

There have been several attempts to eliminate this problem by suitable assumptions about the UV completion of the theory, trying at the same time to understand how sensitive Hawking radiation is to such assumptions~\cite{Brout:1995wp, Corley:1996ar, Barrabes:1998iw, Parentani:1999qv, Barrabes:2000fr, Parentani:2000ts, Parentani:2007mb, Unruh:2004zk}.  It should be noted, however, that the existence of a trans-Planckian problem as defined above does not necessarily imply that trans-Planckian physics affects gravitational collapse and/or the production of Hawking radiation.  In fact, explicit calculations~\cite{Massar:1996tx,Giddings:2006sj} show that the presence of Hawking modes does not alter appreciably the behaviour of collapsing matter, in agreement with the common lore that a freely-falling observer will not detect a substantial amount of radiation even in a neighbourhood of the horizon (see also~\cite{Barcelo:2007yk}).
\vskip 1.5cm
\begin{figure}[htb]
\begin{center}
\includegraphics[width=4cm]{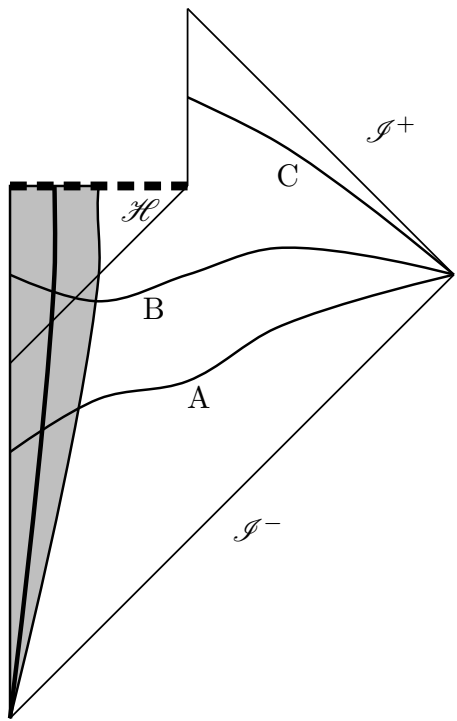}
\end{center}
\vskip 2.3cm
\caption{Conformal diagram describing gravitational collapse followed by Hawking evaporation.  If no leftover remains and Hawking radiation is always thermal, the state on the time slice C is mixed even if it was pure on the time slices A and B.}
\label{figure2}
\end{figure}
%

\subsection{Information loss problem}

The Hawking effect implies that a black hole will eventually evaporate through the emission of thermal radiation.  The present knowledge of fundamental physics does not allow us to predict whether the spacetime associated with such a process is the one illustrated in Fig.~\ref{figure2}, in which black hole evaporation simply leads to a Minkowskian region with diluted thermal radiation flying towards $\scri^+$, or whether other, more complicated final states are allowed.  Among such alternative possibilities, one can consider the evolution of the high-curvature region (classically described as a singularity) into a spatially disconnected baby universe~\cite{ted-valdivia}, the formation of a remnant\footnote{Although logically possible, the remnant hypothesis is somehow disfavoured~\cite{Giddings:1992hh, Giddings:1993km, Giddings:1993vj} (see, however,~\cite{Hossenfelder:2009xq} for a critical assessment).}, and the emission of a final burst of highly correlated non-thermal radiation.  Neglecting these possibilities, \ie, assuming that the process of evaporation is correctly described by the conformal diagram in Fig.~\ref{figure2}, the final quantum state is mixed, even if it was pure at the beginning of collapse.  We are thus in the presence of a  non-unitary evolution~\cite{Hawking:1976ra, Hawking:1982dj}, which is forbidden by the standard rules of quantum theory.   In particular, such an evolution would make it impossible to reconstruct the state of the infalling matter from the emitted radiation.  This is the information loss problem~\cite{Page:1993up, Giddings:1994zs}.  We stress again that the existence of this problem is conditioned by the correctness of the diagram in Fig.~\ref{figure2} and by the exact thermality of Hawking radiation.  Also, the evolution is non-unitary only between time slices like A or B in Fig.~\ref{figure2} {\em and\/} C --- the evolution between A and B is perfectly unitary.

Further explanation is in order, because even in a fully quantum treatment of matter and gravity there could be an apparent difficulty of this sort, if the question about the nature of the final state is not correctly asked.  Imagine that the asymptotic initial state is factorized as $|\Psi_\mathrm{g}\rangle\otimes |\Psi_\mathrm{m}\rangle$, where $|\Psi_\mathrm{g}\rangle$ and $|\Psi_\mathrm{m}\rangle$ are, respectively, the states of the gravitational field and of all the matter fields.  Under a completely unitary evolution the final state is no longer factorized in general, because of the interaction  between matter and gravity.  Tracing upon the gravitational degrees of freedom, the final state of matter fields alone would then be mixed, and would not contain enough information to reconstruct the initial state $|\Psi_\mathrm{m}\rangle$.  However, this would not contradict the postulates of quantum theory, because the missing information would be stored in the correlations between gravity and matter.

Therefore, when asking about the possibility of retrodicting the initial state from the final state, one should consider the {\em total\/} final state, of {\em both\/} matter fields and gravity.  However, the Hawking effect is not limited to matter fields, but entails a thermal flux of gravitons as well.  Hence, the problem of information loss is qualitatively different from the pseudo-problem described above, because even considering the full state of quantum fluctuations of geometry and matter fields together, the evolution is not unitary. 
\subsection{A hidden approximation}
\label{subsec:approx}

The previous presentation of the information loss problem contains actually a hidden approximation.  Hawking radiation is derived assuming that one can work in the covariant quantum gravity scenario~\cite{DeWitt:1967yk, DeWitt:1967ub, DeWitt:1967uc, Veltman:1975vx}, where the metric operator $\hat{\g}_{\mu\nu}$ can be split into a classical background $\g^\mathrm{(c)}_{\mu\nu}$ and small quantum fluctuations $\hat{h}_{\mu\nu}$:
\begin{equation}
\hat{\g}_{\mu\nu} \approx \g^\mathrm{(c)}_{\mu\nu}\,\mathbb{I} +  \hat{h}_{\mu\nu}\;.
\label{g+h}
\end{equation}
Although it is well known that this approach leads to inconsistencies, and that to include higher-order effects one would need the full quantum theory of gravity, one expects it to be an effective description of quantum gravitational effects far from singularities and in regions of low curvature,  independent of the underlying quantum gravity/string theory.  The split~\eqref{g+h} relies on the assumption that the quantum fluctuations of the metric are small.  A  calculation in Minkowski spacetime gives, however,
\begin{equation}
\langle\, \hat{h}_{\mu\nu}(t,{\bf x})\, \hat{h}_{\rho\sigma}(t,{\bf y})\, \rangle \propto \frac{L_\mathrm{P}^2}{({\bf x}-{\bf y})^{2}}\;,
\end{equation}
where $L_\mathrm{P}$ is the Planck length~\cite{Visser:1995cc}.  Thus, at small distances the quantum fluctuations can be overwhelmingly large with respect to the mean field $\g^\mathrm{(c)}_{\mu\nu}$; this is the remark behind the idea of spacetime foam \cite{Wheeler:1957mu, Hawking:1979zw}.  To safely use the covariant quantum gravity scenario, then, one has to introduce a coarse-graining scale larger than $L_\mathrm{P}$, over which the quantum fluctuations $\hat{h}_{\mu\nu}$ are averaged to get rid of the foamy details of spacetime at small scales.  Only in this way, can the condition of small fluctuations be satisfied.

The split~\eqref{g+h} into a classical background field and small quantum fluctuations of the geometry relies therefore on the implicit hypothesis that a coarse-graining of the latter has been performed.  In terms of the quantum gravitational degrees of freedom, this implies that all the states associated with spacetime foam are traced away and made invisible.

\section{A scenario where unitarity is preserved}
\label{sec:scenario}

Attempts to drop unitarity introducing the so-called superscattering operator~\cite{Hawking:1982dj} lead to severe difficulties with energy-momentum conservation~\cite{Banks:1983by}.  Let us therefore investigate whether a conservative resolution of the information loss problem exists, in which the underlying theory obeys the standard quantum rules (so, in particular, it is unitary).  As a general setup, we assume that in the asymptotic past and future there is a macroscopic notion of spacetime, \ie, that the semiclassical approximation to quantum gravity holds.  Furthermore, we assume that in the asymptotic past matter and quantum gravitational fluctuations are described by a pure state, while in the asymptotic future, as a result of collapse and subsequent Hawking evaporation, one is left only with radiation flying towards infinity.  In particular, we exclude the formation of remnants or baby universes.

Of course, in this scenario the state in the asymptotic future must also be pure.  Thus, even accepting that the spectrum of radiation is Planckian, it cannot be fully thermal, and radiation quanta must be correlated.  It is in these correlations that the missing information is stored.  Indeed, the correlations are due to the interaction between Hawking's modes and the infalling matter, so they have a potentiality to encode the state of matter in the far past. 

In order for this resolution to be viable, the relevant interaction must be universal, and such that it cannot be switched off --- gravitation is then the obvious candidate.  This conclusion may seem odd, considering that it is precisely the gravitational backreaction of Hawking radiation that leads to information loss.  However, semiclassical evaporation, based on a classical configuration $(\mathscr{M},\g^\mathrm{(c)}_{\mu\nu})$, is not the whole story.  If the full theory has to be unitary, it is the extra gravitational interaction between the collapsing matter and the outgoing modes, due to the quantum fluctuations $\hat{h}_{\mu\nu}$ (gravitons) around $(\mathscr{M},\g^\mathrm{(c)}_{\mu\nu})$, that generates the correlations needed to preserve information.  Such correlations are encoded into matrix elements of the form\footnote{In principle, all the values of $n$ are needed.  Unfortunately, for $n>2$ these quantities cannot be computed because the theory is non-renormalizable.}
\begin{equation}
\langle \,\mbox{final}\, | \;\mathrm{T}  \left( \int_{\mathscr{M}} \d^{4}x\,\sqrt{-\g^\mathrm{(c)}}\;\hat{T}^{\mu\nu}\,\hat{h}_{\mu\nu}\right)^{n}|\,\mbox{collapsing matter}\,\rangle\;,
\label{eq:transitionamplitude}
\end{equation}
and others involving graviton-graviton scatterings.  Here the T-ordering is taken with respect to a suitable foliation in spacelike hypersurfaces; $|\,\mathrm{final}\,\rangle$ and $|\,\mbox{collapsing matter}\,\rangle$ are the states of late time (on $\scri^+$) outgoing radiation (of matter and gravity) and of early time (on $i^-\cup\scri^-$) infalling matter, respectively.  Whereas the former consists mainly of massless fields, the latter is typically made of massive particles.  We refer to~\cite{Birrell:1982ix}, chapter 9, for a detailed discussion of interacting quantum field theories in curved spacetimes, the definition of transition amplitudes and the apparent ambiguity in the definition of normal ordering with respect to different bases of modes.  

In the following, we will show that this point of view necessarily entangles the trans-Planckian and the information loss problems. 

\section{Gravitational scatterings}
\label{sec:scatterings}

Besides the tensorial structures, the transition amplitudes~\eqref{eq:transitionamplitude} contain integrals of the form
\begin{equation}
\int_{\mathscr{M}} \d^{4}x\, \sqrt{-\g^\mathrm{(c)}}\; f^\mathrm{(in)}_1(x)\cdots f^\mathrm{(in)}_P(x)\, f^\mathrm{(int)}_1(x)\cdots f^\mathrm{(int)}_Q(x) \,f^\mathrm{(out)}_1(x)^\ast\cdots f^\mathrm{(out)}_R(x)^\ast\;,
\label{overlapintegral}
\end{equation}
where $f^\mathrm{(in)}_i$ and $f^\mathrm{(out)}_j$ are incoming and outgoing modes, and $f^\mathrm{(int)}_k$ are virtual modes that appear inside the Feynman diagrams.  Of course, only the region of $\mathscr{M}$ where the modes have a non-negligible overlap contributes to~\eqref{overlapintegral}.  The $f^\mathrm{(in)}_i$ and $f^\mathrm{(out)}_j$ selected by the states in~\eqref{eq:transitionamplitude} can be chosen, respectively, as wavepackets that cross the horizon at very early phases of the black hole life (collapsing matter), and as wavepackets that reach $\scri^+$ at very late times (outgoing radiation).  These functions overlap significantly in a very localized region near the event horizon and in the early phases of evaporation (see Fig.~\ref{figure1}), where semiclassical gravity should be an appropriate description, given that the curvature is very small.\footnote{We consider macroscopic black holes.}  However, in this region the frequency of the outgoing modes is blue-shifted with respect to the asymptotic value on $\scri^+$, and becomes arbitrarily large as one approaches the horizon.  We shall now argue that this leads to a breakdown of the picture.

Since the integral~\eqref{overlapintegral} extends only to a tiny region of small curvature, the amplitude in curved spacetime is indistinguishable from the one evaluated in Minkowski spacetime.  We can then get a feeling for the effect of the gravitational scatterings  between the collapsing matter and the emitted radiation, evaluating the corresponding cross section when the background metric $\g^\mathrm{(c)}_{\mu\nu}$ is flat.  To simplify the algebra, one can perform the calculation in the frame where the massive infalling particle is initially at rest (essentially, a freely-falling frame in the black hole spacetime).  The total cross section diverges, since the interaction is long range.  However, in a partial-wave decomposition, just by dimensional analysis we can conclude that the cross section for each $l$-wave is of the form
\begin{equation}
\sigma_l = c_l\, G_\mathrm{N}^{2}\, s\;,
\label{eq:lthcrosssection}
\end{equation}
where $c_l$ is a numerical coefficient of order one, $G_\mathrm{N}$ is Newton's constant, and $s$ is the Mandelstam variable associated with the centre of mass energy.  This result holds in the high energy limit, when the particle mass is negligible~\cite{DeWitt:1967ub, DeWitt:1967uc}.

By a general theorem of scattering theory, in a unitary evolution $\sigma_l$ is bounded from above.  Its value cannot exceed
\begin{equation}
\sigma_l^\mathrm{max} = b_l/s\;,
\end{equation}
where $b_l$ is again a numerical coefficient of order one.  This is known as the unitarity limit~\cite{perkins}. 

It is now clear that the gravitational scattering at high energies (\ie,  at large $s$) exceeds the unitarity limit.  Apart from dimensionless coefficients of order one, the saturation $\sigma_l=\sigma^\mathrm{max}_l$ occurs at $\bar{s} \sim G_\mathrm{N}^{-1} =M_\mathrm{P}^{2}$.  Given that, in the rest frame of the infalling matter, the frequency of the incoming quantum of radiation is $\omega=s/m$, where $m$ is the particle mass, we see that for quanta with frequencies higher than $\bar{\omega} \sim M_\mathrm{P}^2/m$, the unitarity limit is violated.

This result means that the regime $\omega>\bar{\omega}$ is beyond the domain of applicability of the low energy effective theory we are considering.  In order to cure this problem, one ought to include higher order contributions from the perturbation series. However, this is not possible in our case, because the theory of gravity is perturbatively non-renormalizable.  As in the case of the Fermi theory of weak interactions, for which a similar line of reasoning applies, the crossing of the unitarity limit is a signal that the effective theory is breaking down, and that it must be replaced by a suitable UV completion. 

Strictly speaking, these calculations hold for perturbative quantum gravity around a {\em flat\/} background.  However, we can export the conclusion to our case of a black hole spacetime exploiting the approximate identification we have established between amplitudes in Minkowski and in a curved spacetime.  Although the  corresponding cross sections differ, such a difference is traceable to kinematical features, while the breakdown of the effective field theory is due to an incorrect description of the dynamics, which is encoded into the transition amplitude.

As a consequence, using~\eqref{eq:transitionamplitude} we are going beyond the limits of semiclassical gravity, even outside the event horizon and even in regions where the curvature is very small, where the intuition would suggest that we can trust the semiclassical gravity approximation. 

These conclusions rely on our specific choice for the state $|\,\mathrm{final}\,\rangle$ in Eq.~\eqref{eq:transitionamplitude}, which refers only to the Hawking quanta that fly towards infinity.  This restriction is necessary given that our goal is to reconstruct the state by measurements performed on outgoing radiation only (\ie, on a slice like C in Fig.~\ref{figure2}).  One might consider including in the amplitude the infalling partners as well, and some investigations indeed suggest that this might wash out the effects of the trans-Planckian regime~\cite{Massar:1996tx,Giddings:2006sj}.  However, not only is such a result expected (as we pointed out at the end of Sec.~\ref{subsec:tP}, it supports the common belief that nothing special happens to a freely-falling observer as it crosses the horizon); it is also irrelevant for the issue of information loss, because having access to both members of the Hawking pairs one can obviously know the initial state.  (Alternatively, and equivalently, such calculations only involve the evolution between slices like A and B in Fig.~\ref{figure2}, which is obviously unitary.)  In a more technical vein, it is unclear how this could be implemented in a scattering scheme, because the infalling Hawking quanta cannot be treated as asymptotic states (at least without some ad hoc assumptions about the final outcome of the black hole evaporation).\footnote{It must also be noticed that once one decides to take into account also the infalling Hawking quanta then, by consistency, also gravitational interaction between Hawking partners will have to be taken into account. One could expect then that the state describing pair creation will not be a simple squeezed state as in the test field approximation.}

\section{Linking the problems}
\label{sec:link}

The connection between the information loss and trans-Planckian problems is now clear.  There is a region of spacetime close to the horizon,  where the gravitational interaction between the modes corresponding to Hawking quanta and the infalling matter happens to be in the trans-Planckian regime, thus violating the unitarity limit.  This implies that we cannot trust the covariant quantum gravity approximation close to the black hole horizon, so there is an obstruction to the proposal of information recovery through correlations.

Solving the unitarity limit problem requires introducing higher order corrections into the scattering amplitude, and in the quantum version of general relativity this is impossible because the theory is non-renormalizable.  Of course, no such difficulty would arise in full quantum gravity, but using it might appear an overkill, given that we are dealing with a low-curvature regime, for which a na\"{\i}ve power-counting suggests that the Einstein-Hilbert truncation is well motivated.  One should remember, however, the non-separation between low and high energy physics noticed in section~\ref{subsec:tP}, ultimately traceable to the fact that the unbounded blue-shift between $\scri^+$ and an horizon uncovers the micro-structure of spacetime~\cite{Jacobson:1995ak}.\footnote{The existence of an horizon is, however, not a necessary condition for the existence of the trans-Planckian problem.  For example, the e-folding in inflationary cosmology has basically the same effect as the gravitational redshift in black holes~\cite{Brandenberger:1999sw, Martin:2000xs, Brandenberger:2009rs}.}  Since low curvature does not necessarily imply low energy, the low-curvature split~\eqref{g+h} does not guarantee, in general, that the use of perturbative theory is appropriate.  The crossing of the unitarity limit can then be regarded as a diagnostic tool that signals that the splitting~\eqref{g+h} is no more properly describing the relevant regime.\footnote{The inconsistency of the semiclassical picture of black hole formation and evaporation has also been argued, using both string field theory techniques and arguments about the validity of the effective field theory description, in~\cite{Giddings:2006sj,Kiem:1995iy,Lowe:1995ac,Giddings:2004ud,Giddings:2006be,Giddings:2007ie}.}

In the light of the remarks in section~\ref{subsec:approx} we can, however, say more.  The presence of the trans-Planckian problem implies that any unitary evolution operator describing the Hawking process will have to relate low energy states to ultra-high energy ones.
However, insisting in working on  a background spacetime as in \eqref{g+h},  implies tracing over that part of the Hilbert space corresponding to the microstructure of spacetime below the Planck length. Hence, one should not generically expect that the evolution of low energy states at early times into low energy states at late times will still be unitary.  The trans-Planckian problem is therefore also revealing a basic lack of unitarity of the effective description.  Within the framework in which Hawking radiation is commonly derived (and its extension to covariant quantum gravity), the information loss and trans-Planckian problems can then be reformulated in a unified way as follows:  There is no calculable unitary evolution operator connecting low energy modes on $\scri^+$ with low energy modes on $\scri^-$.
 
Furthermore, while suitable UV completions of the above description might avoid the violation of the unitary bound, they do not generically guarantee an avoidance of the information loss. In fact, any such completion which still involves excitation of trans-Planckian degrees of freedom is exposed to the risk of loosing important correlations (and hence face an information loss problem) once a background spacetime as in Eq.~\eqref{g+h} is introduced via coarse graining.  For example, one could potentially solve the information loss paradox within the renormalization group approach~\cite{Reuter:1996cp,asymptoticsafety,Codello:2008vh}.  (See~\cite{salam} for an earlier suggestion along similar ideas.)  In this case, the inclusion of large quantum fluctuations, hence of the trans-Planckian degrees of freedom, leads to a running of the Newton constant. In particular, if there is a UV fixed point for $G_\mathrm{N}$, at high energies  $G_\mathrm{N}(s) \approx g_\ast/s$, and one can safely use correlations evaluated with gravitational scatterings, provided that the gravitational constant is rescaled.  Therefore, the high energy behavior of the partial wave cross section becomes $\sigma_l \approx c_l\,g_\ast^2/s$, which is compatible with the high-energy behaviour of the cross section imposed by unitarity.\footnote{In~\cite{Bonanno:2006eu} possible implications of the renormalization of gravitational couplings for the global structure of a black hole spacetime have been qualitatively discussed.}  Note however, that this {\em does not} assure in any way that an information loss will not arise, as in principle it does not forbid trans-Planckian degrees of freedom to be involved in the scatterings and to end up encoding some relevant amount of information which will be lost once the coarse grained metric \eqref{g+h} will be introduced.
\section{Final remarks}
\label{sec:conclusions}

In a locally Lorentz-invariant framework, assume that the formation and evaporation of a black hole is described by an effective field theory in which there is a macroscopic notion of background spacetime geometry over which gravitons propagate --- a description that should be reliable in a low energy/small curvature regime.  We have argued that if such a theory contains a trans-Planckian problem, an information loss problem should also be expected.  Indeed, the description adopted involves a coarse-graining over the spacetime foam degrees of freedom.  Since the latter are related to the UV completion of the low energy theory (full quantum gravity), the existence of a trans-Planckian problem implies that they cannot be neglected, \ie, the coarse-graining is not allowed and one should adopt full quantum gravity.  If one nevertheless insists in using the semiclassical description behind~\eqref{g+h} --- \ie, the assumption  that it makes sense to speak about a background spacetime $(\mathscr{M},\g_{\mu\nu}^\mathrm{(c)})$ ---, the absence of these UV degrees of freedom leads to a non-unitary theory, hence to an information loss in black hole evaporation.\footnote{A similar point of view has been advocated in~\cite{Amati:2006fr}.}

Note, that these conclusions apply to any situation in which a trans-Planckian problem, at least as some stage, shows up. Hence, information loss (in the generalized sense of the absence of a calculable unitary operator connecting low energy degrees of freedom on $\scri^-$ to low energy degrees of freedom on $\scri^+$) should still be expected in remnants, regular black holes~\cite{Hayward,Ashtekar}, and quasi-black holes~\cite{Visser:2009xp,Visser:2009pw} scenarios, as they are all based on the ansatz~\eqref{g+h}. 

If perturbative quantum gravity does not work, even in a low curvature regime, which are the scenarios that can potentially solve the information loss problem?  One possibility is that processes involving black hole evaporation can be fully described only using non-perturbative full quantum gravity or string theory, and unitarity is preserved through large quantum fluctuations of the geometry, which is very badly described by a mean classical spacetime geometry.\footnote{However, the derivation of Hawking radiation (which usually assumes the existence of a background spacetime) should also be reconsidered within this framework, asking whether the amplitude $\langle\,\mbox{Hawking radiation}\,|\,\mbox{collapsing matter}\,\rangle$ is non-negligible.}  The AdS/CFT correspondence between a gravitational theory in anti-de Sitter spacetime and a unitary quantum theory, goes in this direction~\cite{Lowe:1999pk, Hawking:2005kf}.

It then seems that the only possibility to avoid information loss within the ansatz~\eqref{g+h}, is that the spacetime foam acts basically just as a catalyzer: While participating in the scattering processes necessary to preserve the correlations, it does not store information, which remains encoded in low energy degrees of freedom.  Indeed, the growing (although not conclusive) evidence of the robustness of the spectrum of Hawking radiation against  the trans-Planckian completion of the theory seems to support this conjecture~\cite{Unruh:1980cg,Unruh:1994je,Brout:1995wp,Corley:1996ar,Barrabes:1998iw,Parentani:1999qv,Barrabes:2000fr,Parentani:2000ts,Parentani:2007mb,Unruh:2004zk}. The detailed description of how this could be the case, would then be a crucial step forward in understanding the UV nature of gravity and spacetime.

\acknowledgments
We would like to thank D. Amati, C. Barcel\'o, S. Hossenfelder, R. Parentani and M. Visser for comments and constructive criticisms on earlier versions of this manuscript. 


\end{document}